\documentclass[12pt, preprint]{aastex}
\usepackage{emulateapj5}
\usepackage{onecolfloat5}
\usepackage{apjfonts}
\usepackage{epsfig}

\def\myputfigure#1#2#3#4#5%
{\vskip#5pt\makebox[0pt]{\hskip#2in
\includegraphics[width=#3\textwidth]{#1}}\vskip#4pt\hfill}

\newcommand\lsim{\mathrel{\rlap{\lower4pt\hbox{\hskip1pt$\sim$}}
        \raise1pt\hbox{$<$}}}
\newcommand\gsim{\mathrel{\rlap{\lower4pt\hbox{\hskip1pt$\sim$}}
        \raise1pt\hbox{$>$}}}

\newcommand{\nf}{x_{\rm HI}}

\newcommand{\strom}{Str\"omgren sphere}
\newcommand{\stromspace}{Str\"omgren sphere~}
\newcommand{\lya}{Lyman~$\alpha~$}
\newcommand{\lyb}{Lyman~$\beta~$}
\newcommand{\taudamp}{\tau_{D}}
\newcommand{\taures}{\tau_{R}}
\newcommand{\lobs}{\lambda_{\rm obs}}
\newcommand{\zsource}{z_{Q}}

\newcommand{\fion}{f_{\rm ion}}
\newcommand{\qname}{SDSS J1030+0524}
\newcommand{\taub}{\tau_{\rm lim(Ly\beta)}}
\newcommand{\taua}{\tau_{\rm lim(Ly\alpha)}}
\newcommand{\tautot}{\tau_{\rm Ly\alpha}}

\begin{document}
\twocolumn[%%% Begin front material
\submitted{Accepted by ApJ Letters}
\title{Evidence for a Cosmological Str\"omgren Surface and for Significant Neutral Hydrogen Surrounding the Quasar SDSS J1030+0524}

\author{Andrei Mesinger, Zolt\'{a}n Haiman}
\affil{Department of Astronomy, Columbia University, 550 West 120th Street, New York, NY 10027}

\begin{abstract}
A bright quasar residing in a dense and largely neutral intergalactic
medium (IGM) at high redshifts ($z\gsim 6$) will be surrounded by a
large cosmological Str\"omgren sphere.  The quasar's spectrum will
then show a sharp increase in resonant Lyman line absorption at
wavelengths approaching and shorter than that corresponding to the
Str\"omgren sphere's boundary along the line of sight.  We show here
that simultaneously considering the measured absorption in two or more
hydrogen Lyman lines can provide the dynamical range required to detect this
feature.  We model broad and robust features of the \lya and \lyb
regions of the spectrum of the $z=6.28$ quasar SDSS J1030+0524, using
a hydrodynamical simulation. From the steep wavelength--dependence of
the inferred absorption opacity, we detect the boundary of the
Str\"omgren sphere at a proper distance of $6.0\pm 0.2$ Mpc away from
the source redshift.  From the spectrum alone, we also find that
beyond this distance, cosmic hydrogen turns nearly neutral, with
a neutral fraction of $x_{\rm HI}\gsim 0.2$, and
that the ionizing luminosity of this quasar is in the range $(5.2\pm
2.5)\times 10^{56}$ photons sec$^{-1}$.  The method presented here,
when applied to future quasars, can probe the complex topology of
overlapping ionized regions, and can be used to study the details of
the reionization process.\\ 
\end{abstract}
]

\section{Introduction}
\label{sec:intro}

The ionization state of the IGM at redshift $6\lsim z \lsim 7$ has
been a subject of intense study over the past few years.  While
studies of the cosmic microwave background anisotropies by the {\it
Wilkinson Microwave Anisotropy Probe (WMAP)} satellite imply that the
IGM is significantly ionized out to $z\sim 15$, several pieces of
evidence suggest that it has a large neutral fraction at $z\sim 6-7$
(see, e.g., Haiman 2003 for a recent review).  If indeed the
intergalactic hydrogen is largely neutral at $z\sim 6$, then
quasars at this redshift should be surrounded by large ionized (HII)
regions \citep{mr00, ch00}.  The imprint on the absorption spectrum of
the damping wing of absorption by neutral hydrogen outside the HII
region \citep{mhc04}, and the size of the HII region itself
\citep{wl04} can be used to place stringent limits on the neutral
fraction of the ambient IGM.

In principle, the quasar's absorption spectrum contains a record of
the neutral fraction as a function of position along the line of
sight.  In Figure~\ref{fig:taus} we illustrate a model for the optical
depth to \lya absorption as a function of wavelength towards a
$\zsource=6.28$ quasar embedded in a neutral medium ($x_{\rm HI}=1$),
but surrounded by a \strom\ with a comoving radius of $R_S = 44$ Mpc.
Around bright quasars, such as those recently discovered \citep{fan01,
fan03} at $z\sim 6$, the proper radius of such Str\"omgren spheres is
expected to be $R_S$ $\approx$ 7.7 $x_{\rm HI}^{-1/3}$ $(\dot N_Q/
6.5\times 10^{57}~{\rm s^{-1}})^{1/3}$ $(t_Q/2\times 10^7~{\rm
yr})^{1/3}$ $[(1+z_Q)/7.28]^{-1}$ Mpc \citep{mr00, ch00}. Here $x_{\rm
HI}$ is the volume averaged neutral fraction of hydrogen outside the
\strom\, and $\dot N_Q$, $t_Q$, and $z_Q$ are the quasar's production
rate of ionizing photons, age, and redshift. The fiducial values are
those estimated for J1030+0524 \citep{hc02, wl04}.  The mock spectrum
shown in Figure 1 was created by computing the \lya opacity from a
hydrodynamical simulation (the procedure is described in \S~2 below).
The optical depth at a given observed wavelength, $\lobs$, can be
written as the sum of contributions from inside ($\taures$) and
outside ($\taudamp$) the \strom, $\tautot = \taures + \taudamp$.  The
residual neutral hydrogen inside the \strom\ at redshift $z<z_Q$
resonantly attenuates the quasar's flux at wavelengths around
$\lambda_\alpha(1+z)$, where $\lambda_\alpha = 1215.67$ \AA\ is the
rest-frame wavelength of the \lya line center. As a result, $\taures$
is a fluctuating function of wavelength (solid curve), reflecting the
density fluctuations in the surrounding gas.  In contrast, the damping
wing of the absorption, $\taudamp$, is a smooth function (dashed
curve), because its value is averaged over many density fluctuations.
As the figure shows, the damping wing of the absorption from the
neutral universe extends into wavelengths $\lobs\gsim 8720$ \AA, and
can add significantly to the total optical depth in this region.

The sharp rise in $\taudamp$ at wavelengths $\lobs \lsim 8720$ \AA\ is
a unique feature of the boundary of the HII region, and corresponds to
absorption of photons redshifting into resonance outside of the
\strom.  The detection of this feature has been regarded as
challenging: since the quasar's flux is attenuated by a factor of
$\exp(-\tautot)$, an exceedingly large dynamical range is required in
the corresponding flux measurements.  {\em However, we show here that
simultaneously considering the measured absorption in two or more
hydrogen Lyman lines can provide the dynamical range required to
detect this feature.} In particular, we model broad features of the
\lya and \lyb regions of the absorption spectrum of the $z=6.28$
quasar SDSS J1030+0524. We find robust evidence for the boundary of
the St\"omgren sphere, and derive new limits on the neutral fraction
of the ambient IGM, and on the ionizing emissivity of the
quasar.\footnote{
Throughout this paper we assume a standard
cold--dark matter cosmology ($\Lambda$CDM), with
($\Omega_\Lambda$, $\Omega_M$, $\Omega_b$, n, $\sigma_8$, $H_0$) =
(0.73, 0.27, 0.044, 1, 0.85, 71 km s$^{-1}$ Mpc$^{-1}$), consistent
with the recent results from {\it WMAP} \citep{spergel03}.  Unless stated otherwise, all lengths are quoted in
comoving units.}

%\clearpage

\vspace{+0\baselineskip}
\myputfigure{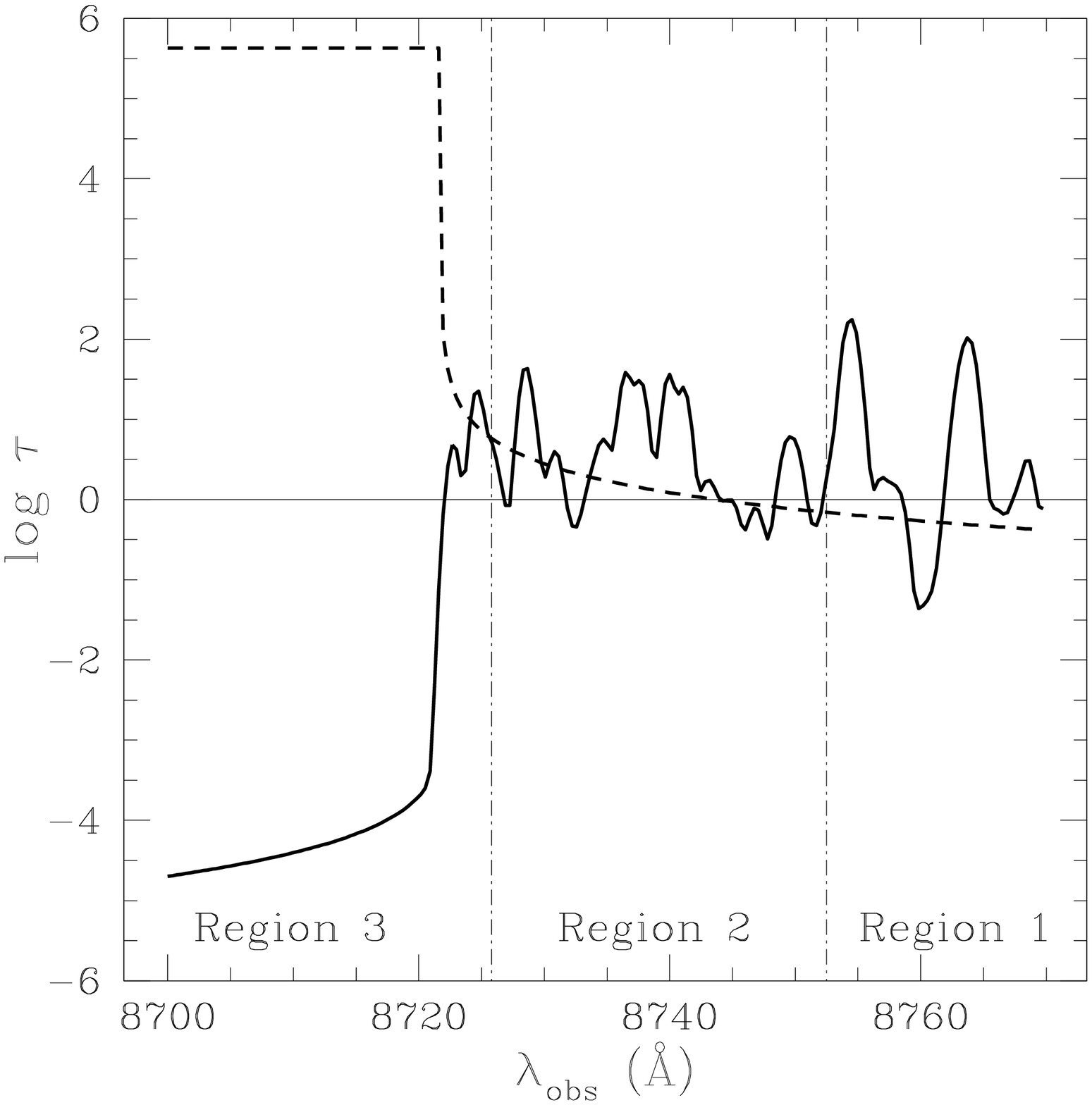}{3.3}{0.5}{.}{0.}  
\vspace{-1\baselineskip} \figcaption{Model from a hydrodynamical
simulation for the optical depth contributions from within ($\tau_R$)
and from outside ($\tau_D$) the local ionized region for a typical
line of sight towards a $\zsource=6.28$ quasar embedded in a fully
neutral, smooth IGM, with $R_S$ = 44 Mpc and $\fion$ = 1.  The
\emph{dashed curve} corresponds to $\taudamp$, and the \emph{solid
curve} corresponds to $\taures$.  The total \lya optical depth is the
sum of these two contributions, $\tautot$ = $\taures$ + $\taudamp$.
The \emph{dashed-dotted lines} demarcate the three wavelength regions
used for our analysis described in the text. In our analysis, the
optical depths are averaged over 3.5 \AA\ wavelength bins, which
decreases the fluctuation of $\taures$. For reference, the redshifted
\lya wavelength is at 8852 \AA, far to the right off the plot. \label{fig:taus}}
\vspace{+0\baselineskip}

%\clearpage

\section{Analysis}
\label{sec:analysis}

The observational input to our analysis is the deepest available
absorption spectrum of \qname\ \citep{white03}.  The
flux detection threshold in the \lya and \lyb regions of this spectrum
correspond to \lya optical depths of $\taua \approx 6.3$ and
$\taub \approx 22.8$ respectively.\footnote{For our purposes, these
optical depths can be taken as rough estimates.  Their precise values
are difficult to calculate, with $\taub$ especially uncertain
\citep{song04,letal02, cm02, fan02}.  However, we have verified that
our conclusions below remain unchanged when the threshold opacities
are varied well in excess of these uncertainties. In particular,
considering ranges as wide as $5.5<\taua<7$ and $10<\taub<30$ would
lead to constraints similar to those we derive below.}  The ratio of
these two numbers is smaller than that of
$(f_\alpha\lambda_\alpha)/(f_\beta\lambda_\beta)$, where $f$ is the
oscillator strength, mainly because of the need to subtract foreground \lya
absorption in the \lyb region.  The Keck ESI spectrum of \qname\
exhibits a strong \lya Gunn-Peterson (GP) trough, with no detectable
flux between wavelengths corresponding to redshifts $5.97 < z < 6.20$,
as well as a somewhat narrower \lyb trough between $5.97< z < 6.18$
\citep{white03}.

To summarize these constraints, we have divided the spectrum into
three regions, shown in Figure~\ref{fig:taus}.  In Region 1, with
$\lobs \geq$ 8752.5 \AA, the detection of flux corresponds to the {\it
upper} limit on the optical depth $\tautot < 6.3$.  Region 2,
extending from 8725.8 \AA\ $\leq \lobs <$ 8752.5 \AA, is inside the
\lya trough, but outside the \lyb trough. Throughout this region, the
data requires 6.3 $\lsim \tautot \lsim 22.8$.  Region 3, with $\lobs
<$ 8725.8 \AA\, has a {\it lower} limit $\tautot \geq$ 22.8.  As
defined, each of these three regions contains approximately eight
pixels.

We modeled the absorption spectrum, attempting to match these gross
observed features. We do not perform a far more demanding
pixel--by--pixel fit to the data, given the relatively poor quality of
the spectra currently available.  We utilized a hydrodynamical
simulation that describes the density distribution surrounding the
source quasar at $z=6.28$.  The details of the simulation are
described elsewhere (Cen 2004, in preparation; Mesinger et
al. 2004). We extracted density and velocity information from 100
randomly chosen lines of sight (LOSs) through the simulation
box.\footnote{The density and velocity distributions are biased near
the density peaks where a quasar may reside. However, the spectral
region of interest lies well outside these biased regions, which, in
the case of high--redshift quasars, extends on average only to $\lsim
1$ proper Mpc\citep{bl04}.  The use of randomly chosen LOSs is
therefore an accurate statistical representation of the expected
density and velocity fields.}  Along each line of sight (LOS), we
computed the \lya absorption as a function of wavelength.  The 
size of the ionized region ($R_S$) and the
fraction of neutral hydrogen outside it ($\nf$) were free parameters.
In addition, the quasar's ionizing luminosity,
$L_{\nu}$, was a third free parameter, $\fion$, defined by $ L_{\nu}
\equiv \fion ~ L^{e}_{\nu}$. Here $L^{e}_{\nu}$ = $1.55 \times
10^{31}$ $(\nu/\nu_H)^{-1.8}$ $[(1+z)/(1+\zsource)]^{-0.8}$,
$\nu_H=3.29\times10^{15}$ Hz is the ionization frequency of hydrogen,
and $\zsource$ is the source redshift ($\zsource$ = 6.28 for \qname).
$L^{e}_{\nu}$ results from redshifting a power--law spectrum with a
slope of $\nu^{-1.8}$, normalized such that the emission rate of
ionizing photons per second is $1.3 \times 10^{57}$, matching the
emissivity one obtains \citep{hc02} for SDSS J1030+0524 using a
standard template spectrum \citep{elvis}. This translates to
an emission rate of ionizing photons of $\dot N_Q\approx\fion\times
1.3\times 10^{57}~{\rm s^{-1}}$.  To understand the significance of
these parameters, note that changing $R_S$ moves the dashed
($\taudamp$) curve in Figure~\ref{fig:taus} left and right, while changing $\nf$ moves it
up and down; changing $\fion$
moves the solid ($\taures$) curve up and down. Values
of $\nf<1$ imply the existence of an ionizing background flux, 
with an ionization rate of $\Gamma_{12}\approx 10^{-5}
x_{\rm HI}^{-1} \times 10^{-12}~{\rm s^{-1}}$.  We
add this background flux to that of the quasar (the latter
dropping as $r^{-2}$ with distance), somewhat reducing the neutral
hydrogen fraction inside the ionized region.

We evaluated $\taures$ and $\taudamp$ for each LOS, in the range of
observed wavelengths $8700$ \AA\ $< \lobs < 8770$ \AA, and smoothed
the optical depth by averaging $\exp(-\tautot)$ over $\sim$ 3.5 \AA\
wavelength bins, corresponding to the Keck ESI spectral resolution of
$R$ = 2500 \citep{white03}. This procedure was repeated for each
combination of $R_S$, $\nf$, and $\fion$. We allowed $\nf$ to range
from 1 to $10^{-3}$, to trace its currently allowed range
\citep{fan02}.  Values of $R_S <$ 42 Mpc (5.8 Mpc proper) are
immediately ruled out by the presence of flux at $z\gsim 6.18$ in the
redshifted \lyb region of the spectrum. Hence, we let $R_S$ range from
42 to 54 Mpc, in increments of 1 Mpc.  Finally,
$\fion$, was varied from 0.1 to 10, in increments of 0.1.  For each
point in the three--dimensional parameter space of $R_S$, $x_{\rm
HI}$, and $\fion$, we computed the fraction of the 100 LOSs that were
acceptable descriptions of the spectrum of \qname, based on the
criteria defined above.  To be conservative, we
allowed up to two pixels in each region to lie outside the allowed
range of \lya optical depths; each LOS that had more than two pixels
fail the above criteria in any of the three regions was rejected. We
find that a more stringent criterion of allowing only a single
'faulty' pixel would strengthen our conclusions.

\vspace{1\baselineskip}
\myputfigure{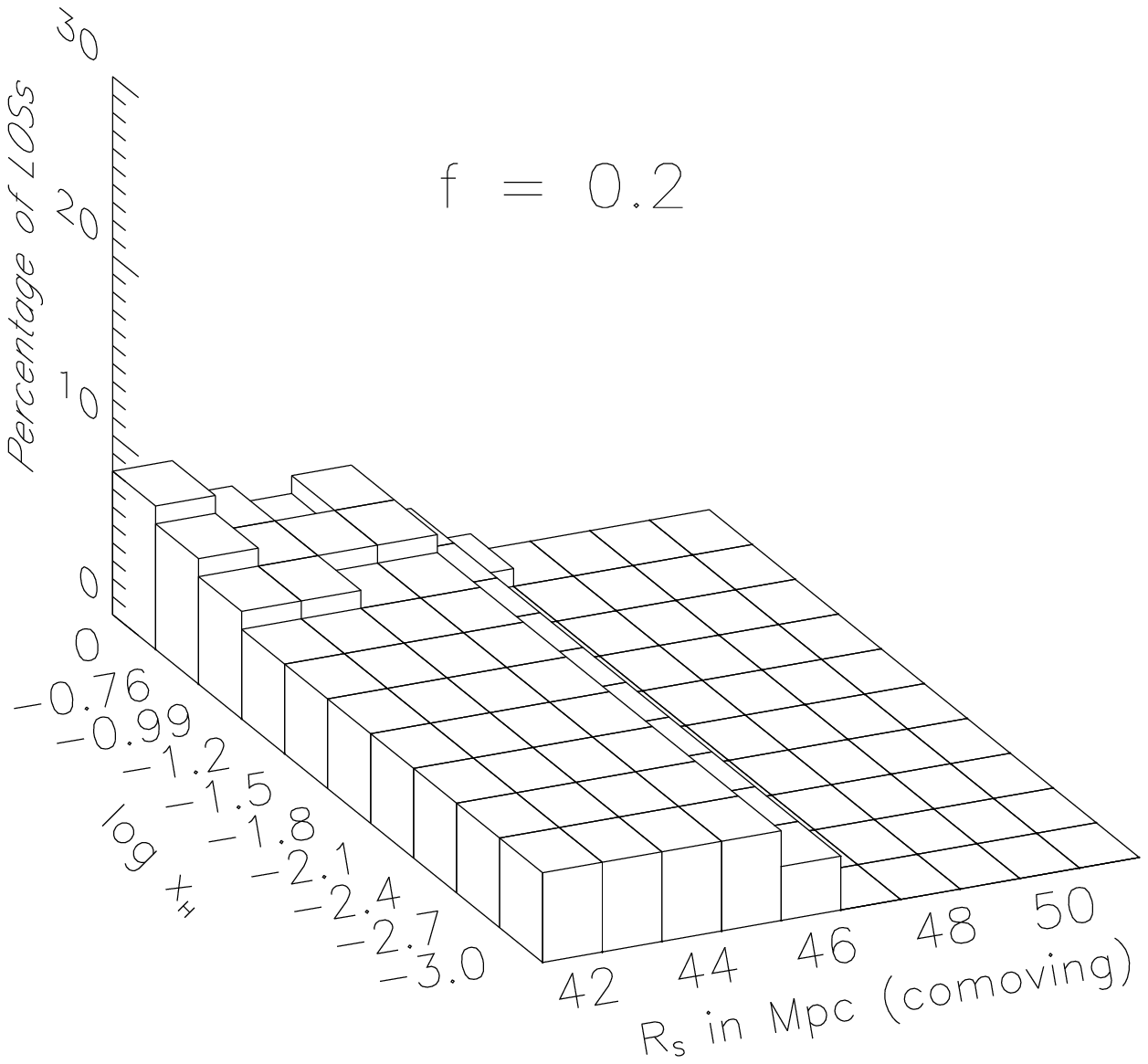}{3.3}{0.4}{.}{0.}  
\vspace{-2\baselineskip}
\myputfigure{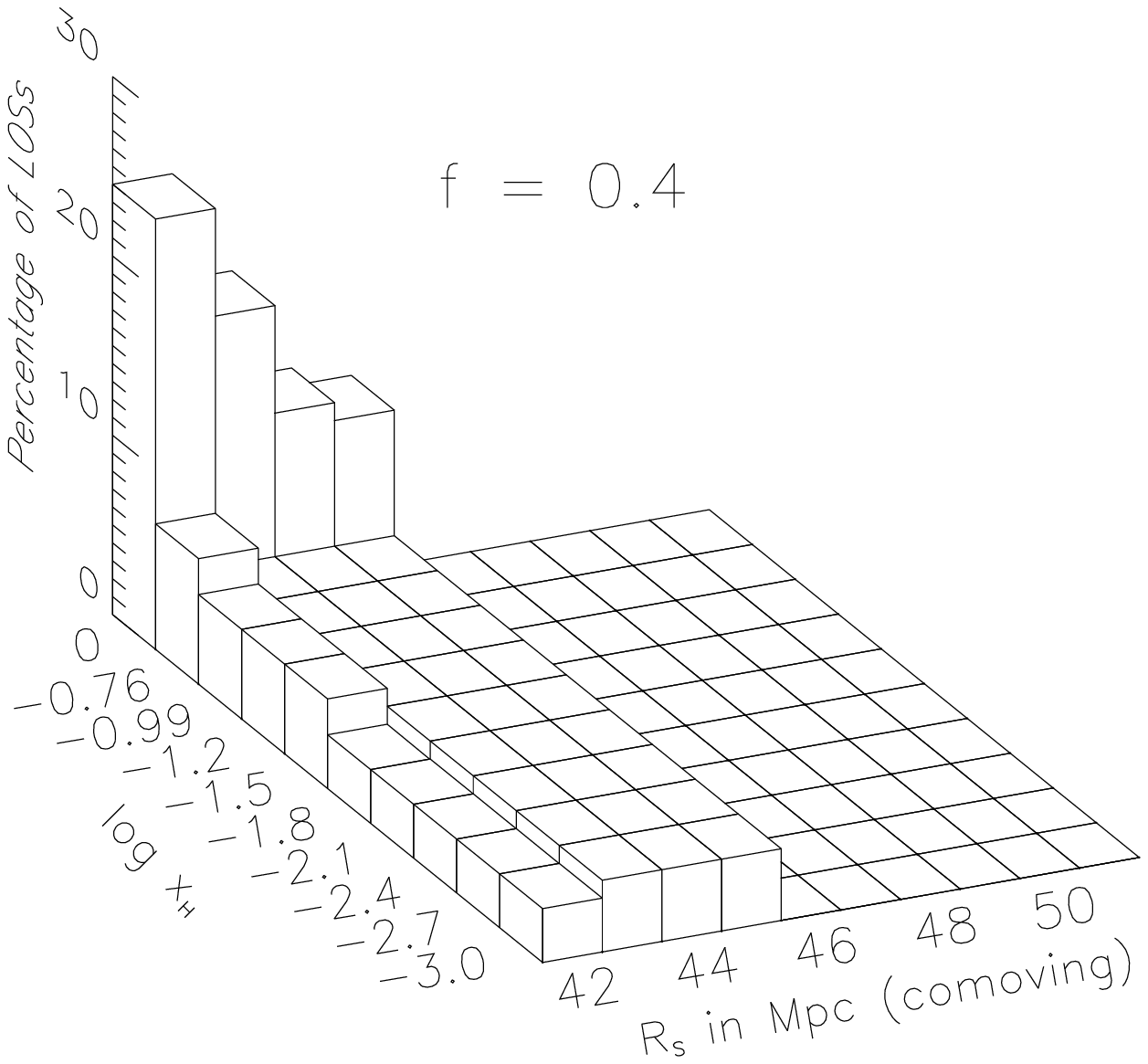}{3.3}{0.4}{.}{0.}
\vspace{-2\baselineskip}
\myputfigure{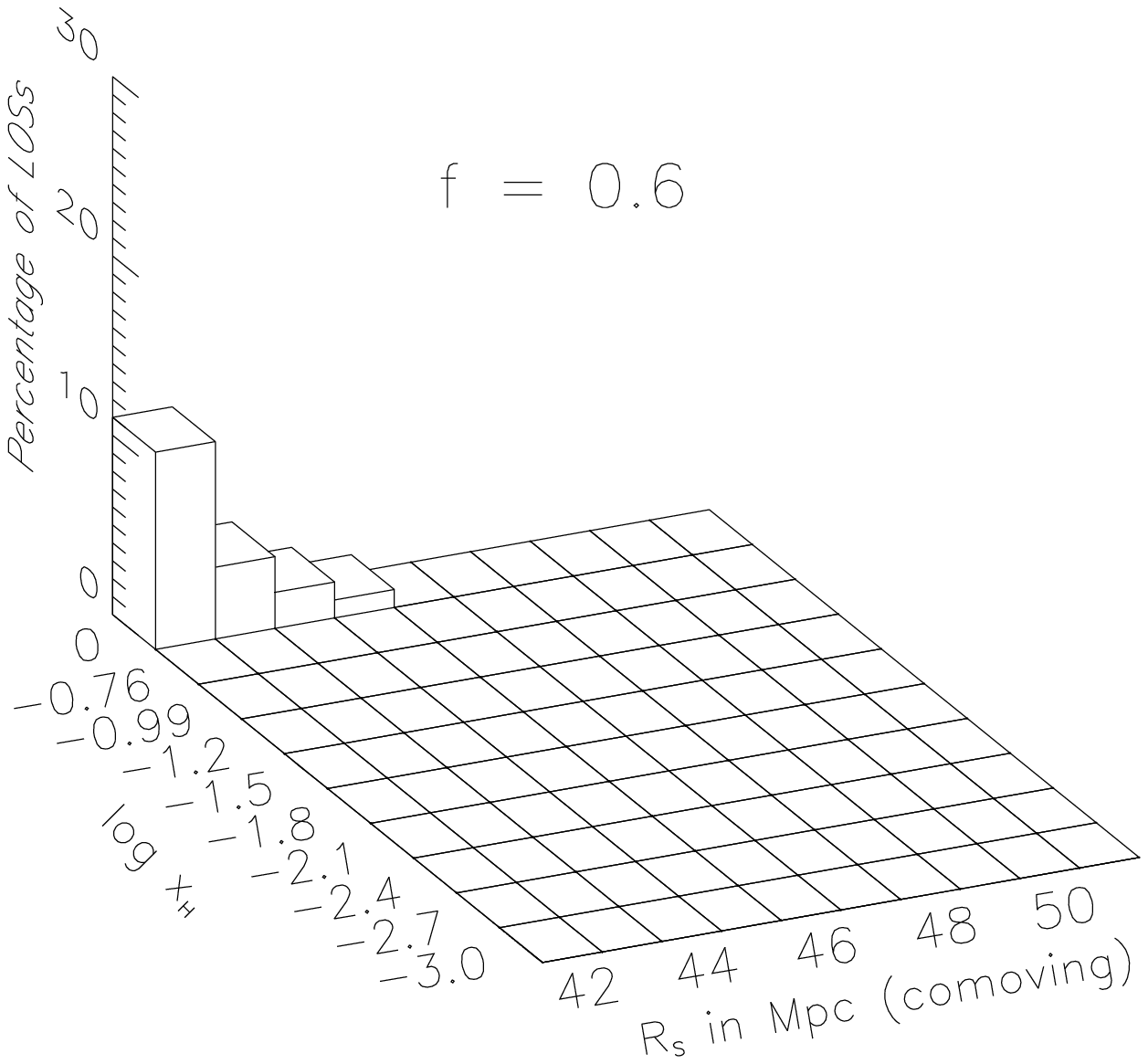}{3.3}{0.4}{.}{0.}
\vspace{0\baselineskip} \figcaption{The percentages of LOSs that
pass our analysis criteria.  Results for three different values of the
ionizing flux parameter, $\fion$ = 0.2, 0.4, 0.6 are shown in the
\emph{top}, \emph{middle} and \emph{bottom} panels, respectively. The
fraction of accepted LOSs peaks  sharply at a maximum of
24\% at the parameter choice of ($R_S$, $\nf$, $\fion$) = (42 Mpc, 1,
0.4).  The radius of the ionized region surrounding the quasar is
limited to 42 Mpc $\leq R_S \leq$ 47 Mpc.  The ionization flux
parameter is tightly constrained around $\fion = 0.4$, with no LOSs
passing our criteria with values of $\fion < 0.2$ and $\fion > 1.0$.
The peak at $\nf = 1$ is over 3.4 times as likely as the next closest
value of $\nf = 0.17$, in which 7\% of the LOSs
pass.\label{fig:vary_f} }
%\vspace{1\baselineskip}

\section{Results}
\label{sec:results}

The procedure outlined above turns out to provide tight constrains on
all three of our free parameters {\it simultaneously}.  The results
are summarized in Figure~\ref{fig:vary_f}, showing the percentage
of LOSs that pass our analysis criteria, as a function of $R_S$,
$\nf$, and $\fion$. The fraction of accepted LOSs peaks sharply at a
maximum of 24\% at the parameter choice of ($R_S$, $\nf$, $\fion$) =
(42 Mpc, 1, 0.4).  The radius of the ionized sphere is limited to 42
Mpc $\leq R_S \leq$ 47 Mpc, with no LOSs passing our criteria outside
of that range.  Note that an offset in the \lya line center by $\pm
1000~{\rm km~s^{-1}}$ (typical of quasar jets, for example), would
represent an additional $\pm 1.5$ Mpc uncertainty on the inferred
radius. Although \qname\ has a precise redshift determination from C
IV and N V lines, a redshift error of $\Delta z =0.01-0.02$ (typical
in the absence of such metal lines) would add an uncertainty of $\pm
(4-8)$ Mpc to any analysis similar to the one presented here.  The
ionization flux parameter is tightly constrained around $\fion=0.4$,
with no LOSs passing our criteria with values of $\fion < 0.2$ and
$\fion > 1.0$.  The peak value of $\fion=0.4$ implies an emission rate
of $(5.2\pm 2.5)\times 10^{56}~{\rm photons~sec^{-1}}$.  Finally, the
peak at $\nf = 1$ is over 3.4 times as likely as the next closest
value of $\nf = 0.17$, in which 7\% of the LOSs pass, and 6 times as
likely as values of $\nf < 0.016$, in which 4\% of the LOSs pass.

These results can be interpreted as follows.  As mentioned previously,
the presence of flux in Region 2 ($\tautot <$ 22.8) sets a {\it lower
limit} on $R_S$, implying that the edge of the ionized region must be
at a smaller wavelength than the boundary between Regions 2 and 3.
Region 3, however, yields an {\it upper limit} on $R_S$, from the
requirement that $\tautot >$ 22.8 in that region. This high optical
depth cannot be maintained by $\taures$ alone, without violating the
constraint in Region 1 of $\tautot <$ 6.3. We emphasize that the {\it
upper limit} on $R_S$ makes no use of the data in Region 2.  Hence the
edge of the ionized region must be close to the boundary between
Regions 2 and 3.  The ionization flux is constrained due to our
criteria in Regions 1 and 2.  An ionization flux that is too small
({\it large} $\taures$) would violate $\tautot <$ 6.3 in Region 1.  An
ionization flux that is too large ({\it small} $\taures$) would
violate $\tautot >$ 6.3 throughout Region 2, since the damping wing
contribution to $\tautot$ is insufficient, given the lower limit on
$R_S$ above.  Our constraint on the neutral hydrogen fraction, $\nf$,
comes from the presence of flux in the \lyb region of the spectrum
corresponding to Region 2. Because of fluctuations in the density
field (and hence in $\taures$), a strong damping wing is needed to
raise $\tautot$ above 6.3 throughout Region 2, {\it while still
preserving} $\tautot < 6.3$ in Region 1.

%\clearpage

%\clearpage

\section{Discussion and Conclusions}
\label{sec:conclusion}

Using only the lower limits on the \lya optical depth obtained from
the \lya and \lyb GP troughs of \qname, we find a simultaneous
constraint on the size of the ionized region surrounding this quasar,
the neutral fraction of the ambient intergalactic medium, and the
ionizing luminosity of the quasar.

The radius of the \stromspace is limited to 42 Mpc $\leq R_S \leq$ 47
Mpc, close to the previously estimated lower limits \citep{mr00,
ch00}.  We also find evidence that the IGM was significantly neutral
at $z \sim 6$, with a $\sim 1$ $\sigma$ lower limit of $\nf\gsim
0.17$.  This result is derived from the observed sharpness of the
boundary of the HII region alone, and relies only on the gross density
fluctuation statistics from the numerical simulation.  In particular,
it does not rely on assumptions about the mechanism for the growth of
the HII region.  Our results provide a robust confirmation of previous
suggestions that the IGM was significantly more neutral at $z\sim 6$
than the lower limit of $x_{\rm HI}\gsim 10^{-3}$ that is directly
obtainable from the black region (Gunn--Peterson trough) of the quasar
spectra.  Finally, we find an emission rate of ionizing photons per
second of $(5.2\pm 2.5)\times 10^{56}~{\rm photons~sec^{-1}}$, which
is between 2 -- 10 times lower than expected \citep{elvis,telfer},
strengthening arguments that reionization at $z\sim 6$ is caused by
the radiation from early stars, rather than bright quasars
(e.g. Dijkstra et al. 2004).

Our findings represent the first detection of the boundary of a
cosmological HII region, and have several important implications.  It
has recently been shown that tight constraints on the hydrogen neutral
fraction can be extracted directly from the \lya absorption spectrum
alone \citep{mhc04}, but this requires tens of spectra to be
statistically significant when $R_S$ is not known.  Incorporating an
independent limit on $R_S$, such as the one obtained from this method,
can reduce the number of required spectra to {\it one}.  Tight
constraints on $\nf$ have also recently been obtained by adopting a
proper size of $R_S\approx 4.5$ Mpc for \qname, together with an
estimate of its lifetime \citep{wl04}. Our direct determination of the
\strom\ size is only slightly larger than the previously adopted
value, lending further credibility to this conclusion.

The sharp boundary we detect also constrains the hardness of the
ionizing spectrum of \qname.  We infer here a rise in the neutral
fraction over a redshift interval of $\Delta z\sim 0.02$,
corresponding to a proper distance of $\sim 1.2$ Mpc. In order for its
mean free path not to significantly exceed this thickness, the energy of
the typical ionizing photon emitted by \qname~must be $<230$eV.  This 
implies that the effective slope of the ionizing
spectrum of this source is softer than $-d\ln L_\nu/
d\ln \nu=1.07$.  This analysis ignores radiative transfer 
effects, which can further blur the apparent boundary of the HII 
region, and would strengthen this limit.

The analysis presented here uses only the \lyb and \lya absorption
spectra for the single quasar J1030+0524, but can be extended by
incorporating additional Lyman lines, which provide further
constraints on the optical depth.  Although such analysis will have
additional uncertainties associated with the need to subtract
foreground absorption due to the lower Lyman transitions, it is likely
that adding the Lyman~$\gamma$ line can tighten the constraints
obtained here \citep{fan02}.  Note that the presence of dust within the 
HII region can attenuate and redden the observed spectrum. This will
not, however, impact our conclusions, since the overall attenuation
is absorbed into $\fion$, and reddening extends over a wavelength
range much broader than the sharp features considered here.

In the future, given a larger sample of quasars (and/or a sample of
gamma--ray burst afterglows with near-IR spectra) at $z>6$, it will be
possible to use the method presented here to search for sharp features
in the absorption spectrum from intervening HII regions, not
associated with the background source itself.  We plan study the
utility of this approach in a future paper. For such external HII
regions that happen to intersect the line of sight, both the red and
the blue sides of each GP trough can, in principle, be detected, and
can yield two separate measurements of the ionized fraction at
different points along the line of sight.  The Universe must have gone
through a transition epoch when HII regions, driven into the IGM by
quasars and galaxies, partially percolated and filled a significant
fraction of the volume.  The detection of the associated sharp
features in future quasar absorption spectra will provide a direct
probe of the the 3D topology of ionized regions during this crucial
transition epoch.  Deep surveys equipped with sufficiently red filters
(with instruments such as those being carried out by the VLT, and
ultimately with deep surveys covering a significant portion of the
sky, such as the survey proposed with the Large-aperture Synoptic
Survey Telescope, LSST; \citet{tt03}) will be able to deliver a large
sample of bright $z>6$ quasars needed for such studies.

\acknowledgments{The authors thank R. Cen for permitting the use
of his simulation and R. White for providing the spectrum of \qname.
This work was supported in part by NSF through grants AST-0307200 and
AST-0307291 and by NASA through grant NAG5-26029.}

\end{document}